# Performance of CRC Concatenated Pre-transformed RM-Polar Codes

Bin Li, Jiaqi Gu, Huazi Zhang
Wireless Technology Lab, Huawei Technologies, P. R. China
{binli, gujiaqi, zhanghuazi}@huawei.com

*Abstract*— **In this paper, we discuss pre-transformed RM-Polar codes and cyclic redundancy check (CRC) concatenated pre-transformed RM-Polar codes. The simulation results show that the pre-transformed RM-Polar (256, 128+9) code concatenated with 9-bit CRC can achieve a frame error rate (FER) of $10^{-3}$ at $E_b/N_o = 1.95dB$ under the list decoding, which is about 0.05dB from the RCU bound. The pre-transformed RM-Polar (512, 256+9) concatenated with 9-bit CRC can achieve a FER of $10^{-3}$ at $E_b/N_o = 1.6dB$ under the list decoding, which is 0.15dB from the RCU bound.**

## I. INTRODUCTION

Polar codes are a major breakthrough in coding theory [1]. They can achieve Shannon capacity with a simple encoder and a simple successive cancellation decoder when the code block size is large enough. But for moderate lengths, the error rate performance of polar codes with the SC decoding is not as good as LDPC or turbo codes. A SC-list decoding algorithm was proposed for polar codes [2], which performs better than the simple SC decoder and performs almost the same as the optimal ML (maximum likelihood) decoding at high SNR. In order to improve the minimum distance of polar codes, either RM-Polar codes [3], or the concatenation of polar codes with CRC [4] and PC [5] were proposed to significantly enhance error-rate performance. Recently, a new PAC (polarization-adjusted convolutional) code was proposed [6], by performing a convolution operation before RM (128,64) code, the PAC (128, 64) code can provide a much better error-rate performance than RM (128,64) code. This is because that a pre-transformation with an upper-triangular matrix (including convolution matrix) does not reduce the code minimum distance but can reduce the number of codewords with the minimum distance [7]. In this paper, we discuss pre-transformed RM-polar codes and CRC concatenated pre-transformed RM-polar codes. The simulations show that the CRC-concatenated pre-transformed RM-Polar codes can approach the RCU bound for (256,128) and (512,256) codes.

In section II, we introduce the encoding of pre-transformed RM-Polar codes and CRC-concatenated pre-transformed RM-Polar codes and in section III we provide some simulation results and compare different codes of (256,128) and (512,256). In Section IV, we draw some conclusions.

## II. PRE-TRANSFORMED RM-POLAR CODES

### A. Encoding of Pre-Transformed Polar/RM-Polar Codes

$F = \begin{bmatrix} 1 & 0 \\ 1 & 1 \end{bmatrix}$, $F^{\otimes n}$ is a $N \times N$ matrix, where $N = 2^n$, $\otimes n$ denotes $n$th Kronecker power, and $F^{\otimes n} = F \otimes F^{\otimes (n-1)}$. Let $H_N = F^{\otimes n}$, the pre-transformed Polar/RM-Polar codes can be generated as

$$X = U \times T \times H_N \qquad (1)$$

where $T$ is an upper-triangular matrix with elements: $T_{i,j} = 0$, if $i > j$; $T_{i,j} = 1$, if $i = j$; $T_{i,j} \in \{0,1\}$, if $i < j$. $U = (u_1, u_2, \dots, u_N)$ is the encoding bit sequence. According to the principle of Polar design, these encoding bits $(u_1, u_2, \dots, u_N)$ have different reliabilities, and these $N$ bits are divided into two subsets according to their reliabilities. The top $K$ most reliable bits are used to send information and the rest are frozen bits set to zeros. But for the RM-Polar codes, both the bit reliability and its weight (the number of ones of its corresponding rows in matrix $H_N$) are considered. The top K most reliable bits are selected among the bits with weights larger than a pre-defined threshold. The RM-Polar codes usually have larger minimum distance than Polar codes and therefore performs better than Polar codes under the list decoding. Since the pre-transformation does not reduce the minimum distance of RM-Polar codes and can reduce the number of codewords with the minimum distance, pre-transformed RM-Polar can performs better than the RM-Polar codes.

In order to further increase the minimum distance or reduce the number of codewords of the minimum distance, the concatenation of pre-transformed RM-Polar codes with CRC is a simple solution. The encoding bits $U$ is composed of information bits and CRC bits as $U = [U_I \; U_{crc}]$, where $U_I$ is information bit vector and $U_{crc}$ is CRC bit vector.

## III. PEROFRMANCE RESULTS

### A. Pre-Transformed RM-Polar (256,128) Code

In this sub-section, we provide the simulation results for Polar code (256,128), RM-Polar (256,128), pre-transformed (PT) RM-Polar (256,128), and CRC-concatenated PT-RM-Polar (256,128).

The Table I shows the minimum distance and the number of codewords of the minimum distance of Polar code (256,128), RM-Polar code (256,128) and PT-RM-Polar code (256,128), respectively, where the PT-RM-Polar code uses a random upper-triangular matrix, and $d_{min}$ and $N_{min}$ are obtained by list decoding with very large list size under very high SNR[8]. It is shown that both RM-Polar and PT-RM-Polar have larger minimum distance than Polar code. The PT-RM-Polar has less number of codewords of minimum distance than RM-Polar code. Fig. 1 shows FER performance of Polar code, RM-Polar and PT-RM-Polar under the list decoding. It is shown that the Polar and RM-Polar codes can achieve the ML bounds with the list size $L$=32 while the PT-RM-Polar code can achieve the ML lower bound with $L$=64. As expected the PT-RM-Polar code performs the best due to the best $d_{min}$ and $N_{min}$.

The CRC concatenated PT-RM-Polar code (256,128) is generated by 1) appending $K_{crc}$ CRC bits to 128 information bits generating 128+$K_{crc}$ bits; 2) encoding these 128+$K_{crc}$ bits by PT-RM-Polar code (256,128+$K_{crc}$). The Table II shows the number of codewords ($N_{min}$) of the minimum distance of the PT-RM-Polar code (256, 128 + $K_{crc}$), and the number of codewords ($N_{min}^*$) of the minimum distance of the CRC-PT-RM-Polar code (256,128). For example, when $K_{crc}$=6, the PT-RM-Polar code (256,128+6) has the number of codewords with the minimum distance $N_{min} = 13376$, among these codewords, there are 213 codewords that can pass the 6-bit CRC. Therefore the 6-bit-PT-RM-Polar code has the number of codewords of the minimum distance $N_{min}^* = 213$. Fig. 2 shows FER performance of 3/6/9-bit CRC-concatenated PT-RM-Polar code under the list decoding. The more the number of CRC bits, the less the number of codewords of minimum distance, and the better the ML performance. It is shown that the 6-bit-CRC-PT-RM-Polar code can achieve the ML bound with the list size $L$=2048 and is about 0.1dB away from the RCU bound at FER of $10^{-3}$, and the 9-bit-CRC-PT-RM-Polar code can achieve the ML bound with very large list size $L$=16384 and is about 0.05dB away from the RCU bound at FER of $10^{-3}$. Because the CRC-PT-RM-Polar codes have much less number of codewords of the minimum distance than the RM-Polar/PT-RM-Polar codes, their ML performance is much better than that of the RM-Polar/PT-RM-Polar codes. Fig.3 shows the FER performance of 6-bit-PT-RM-Polar code under the list decoding with different list sizes.

TABLE I. $d_{min}$ and $N_{min}$ for different (256,128) Codes.

|  | Polar (256,128) | RM-Polar (256,128) | PT-RM-Polar (256,128) |
|---|---|---|---|
| $d_{min}$ | 8 | 16 | 16 |
| $N_{min}$ | 96 | 54576 | 13472 |

TABLE II. $N_{min}$ for CRC-PT-RM-Polar Codes (256,128+$K_{crc}$)

| $K_{crc}$ | 3 | 5 | 6 | 7 | 8 | 9 |
|---|---|---|---|---|---|---|
| $d_{min}$ | 16 | 16 | 16 | 16 | 16 | 16 |
| $N_{min}$ | 14144 | 13600 | 13376 | 13848 | 13888 | 14776 |
| $N_{min}^*$ | 1834 | 438 | 213 | 116 | 56 | 24 |

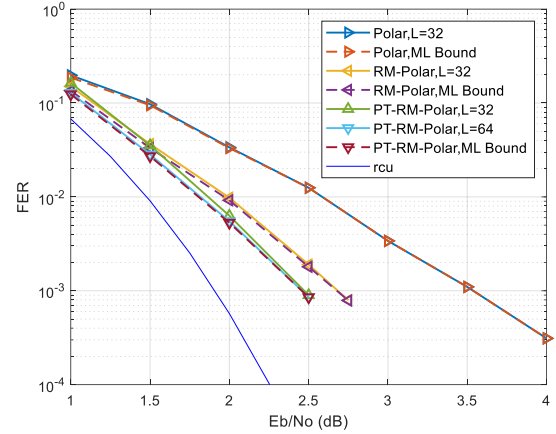

Fig. 1. FER performance of Polar/RM-Polar/PT-RM-Polar (256,128) codes.

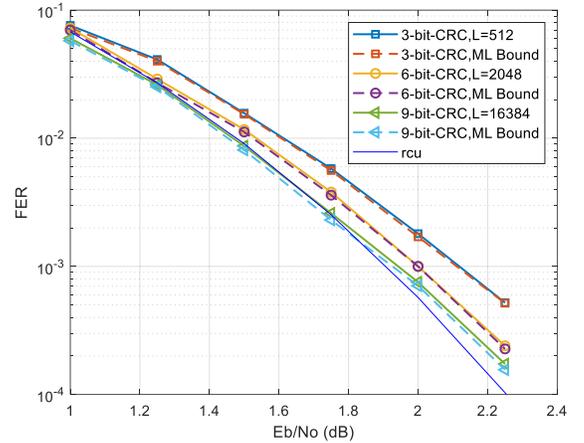

Fig. 2. FER performance of 3/6/9-bit-CRC-PT-RM-Polar (256,128) code.

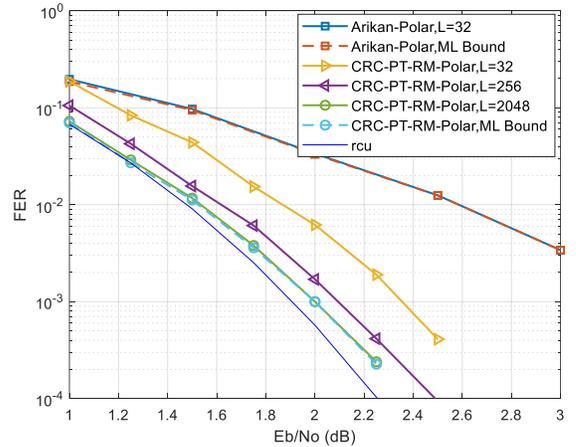

Fig. 3. FER performance of 6-bit-CRC-PT-RM-Polar (256,128) code.

## B. Pre-Transformed RM-Polar (512,256) Code

In this sub-section, we provide the simulation results for Polar code (512,256), RM-Polar (512,256) and PT-RM-Polar (512,256), and CRC-concatenated PT-RM-Polar (512,256).

The Table III shows the minimum distance and the number of codewords of the minimum distance of Polar code, RM-Polar code and PT-RM-Polar code, respectively, where the PT-RM-Polar code uses a random upper-triangular matrix. It is shown that both RM-Polar and PT-RM-Polar have larger minimum distance than Polar code. The PT-RM-Polar has less number of codewords of minimum distance than RM-Polar code. Fig. 4 shows FER performance of Polar code (512,256), RM-Polar (512,256) and PT-RM-Polar (512,256) under the list decoding. It is shown that the list size $L$=32 can achieve the ML lower bound for all three codes. As expected the PT-RM-Polar code performs the best due to the best $d_{min}$ and $N_{min}$.

Fig. 5 shows FER performance of 6/9-bit CRC-concatenated PT-RM-Polar code under the list decoding. The 9-bit-CRC-PT-RM-Polar code (512,256) can achieve FER of $10^{-3}$ at $E_b/N_o = 1.6 dB$, which is about 0.15dB away from the RCU bound. It is interesting to compare CRC-concatenated PT-RM-Polar with CRC-concatenated Arikan-Polar as shown in Fig 6. It is shown that under the $L$=32/$L$=1024, CRC-concatenated PT-RM-Polar performs much better than CRC-concatenated Arikan-Polar.

TABLE III. $d_{min}$ and $N_{min}$ for Different (512,256) Codes.

|  | Polar (512,256) | RM-Polar (512,256) | PT-RM-Polar (512,256) |
|---|---|---|---|
| $d_{min}$ | 8 | 16 | 16 |
| $N_{min}$ | 64 | 63072 | 36544 |

## IV. CONCLUSIONS AND COMMENTS

In this paper, we provide some simulation results for Polar/RM-Polar/PT-RM-Polar/CRC-PT-RM-Polar codes for (256,128) and (512,256), respectively. It is shown that CRC-PT-RM-Polar codes can approach RCU bounds for these two codes. The CRC-PT-RM-Polar code in this paper uses a random pre-transformation matrix. It is unknown how to jointly optimize the CRC bits and the pre-transformation matrix.

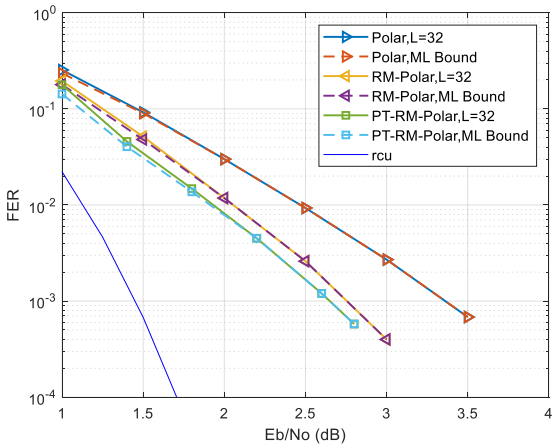

Fig. 4. FER performance of Polar/RM-Polar/PT-RM-Polar (512,256) codes.

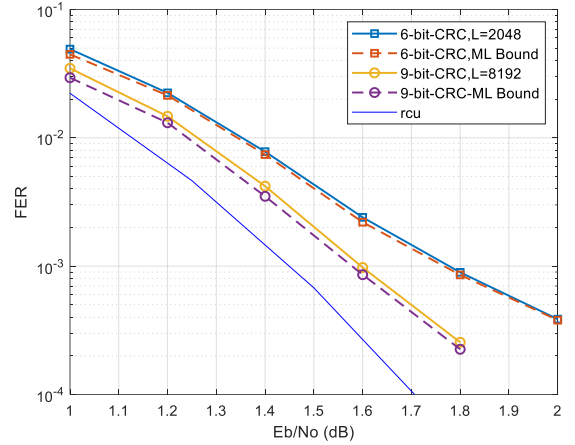

Fig. 5. FER performance of 6/9-bit-CRC-PT-RM-Polar (512,256) codes.

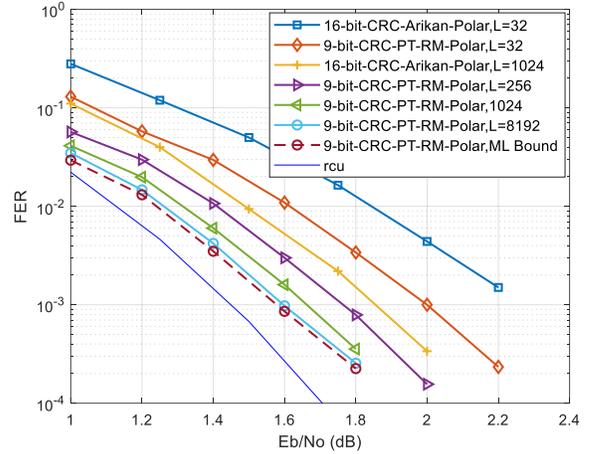

Fig. 6. FER performance of 9-bit-CRC-PT-RM-Polar (512,256) codes.